FRONT MATTER

**Title**

- **Cosmic-Ray Bath in a Past Supernova Gives Birth to Earth-Like Planets**

DOI: https://www.science.org/doi/10.1126/sciadv.adx7892

**Authors**

Ryo Sawada[1,2]*, Hiroyuki Kurokawa[2,3], Yudai Suwa[2,4], Tetsuo Taki[2], Shiu-Hang Lee[5,6], Ataru Tanikawa[7].

**Affiliations**

[1] Institute for Cosmic Ray Research, The University of Tokyo, Chiba, 277-8582, Japan.

[2] Department of Earth Science and Astronomy, The University of Tokyo, Tokyo, 153-8902, Japan.

[3] Department of Earth and Planetary Science, The University of Tokyo, Tokyo, 113-0033, Japan.

[4] Yukawa Institute for Theoretical Physics, Kyoto University, Kyoto, 606-8502, Japan.

[5] Department of Astronomy, Kyoto University, Kyoto, 606-8502, Japan.

[6] Kavli Institute for the Physics and Mathematics of the Universe, The University of Tokyo, Chaba, 277-8583, Japan.

[7] Center for Information Science, Fukui Prefectural University, Fukui, 910-1142, Japan.

*Corresponding author. Email: ryo@g.ecc.u-tokyo.ac.jp

**Abstract**

A key question in astronomy is how ubiquitous Earth-like rocky planets are. The formation of terrestrial planets in our solar system was strongly influenced by the radioactive decay heat of short-lived radionuclides (SLRs), particularly $^{26}$Al, likely delivered from nearby supernovae. However, current models struggle to reproduce the abundance of SLRs inferred from meteorite analysis without destroying the protosolar disk. We propose the 'immersion' mechanism, where cosmic-ray nucleosynthesis in a supernova shockwave reproduces estimated SLR abundances at a supernova distance (~1 pc), preserving the disk. We estimate that solar-mass stars in star clusters typically experience at least one such supernova within 1 pc, supporting the feasibility of this scenario. This suggests solar-system-like SLR abundances and terrestrial planet formation are more common than previously thought.

# MAIN TEXT

## Introduction

The prevalence of Earth-like planets is one of our fundamental questions to the universe. Earth possesses a small but certain amount of liquid water, which allows the atmosphere-ocean-crust interactions and characterizes its habitable environment. It has been proposed that the desiccation of planetesimals is crucial for forming water-depleted rocky planets (a bulk mass fraction lower than 1%) such as Earth (*1,2*). Parent bodies of differentiated meteorites in the solar system are known to have experienced substantial heating due to radioactive decay of a short-lived radionuclide (SLR) $^{26}$Al (*3*), and lost originally-accreted water and other volatiles (*2, 4*). Meteorite analysis has found that SLRs with half-lives shorter than 5 million years ($^{10}$Be, $^{26}$Al, $^{36}$Cl, $^{41}$Ca, $^{53}$Mn, and $^{60}$Fe) existed globally in the early solar system (*5*). In contrast, SLR-depleted systems, if exist, may only form ocean planets whose bulk water content is a few tens of percent (*6*). Therefore, understanding the origin of SLRs in the solar system is crucial to answering the above-mentioned question of the prevalence of Earth-like planets in other stellar systems.

### *Origin of SLRs*

The excess of SLRs in the early solar system provides critical insight into the formation of the solar system (see **Supplementary Text** for more details). The abundance of SLRs in the 'initial solar system' is derived from meteorite analysis (see in Table 1) (*5*), inferring their levels at the formation of Ca-rich, Al-rich Inclusions (CAIs), the first solids in the solar system (*7*). Given their short half-lives, the estimated SLR abundances are too high to have been inherited solely from the parent molecular cloud prior to the onset of solar system formation (*8–10*). Moreover, inheritance from the molecular cloud alone cannot account for the coexistence of $^{26}$Al-rich and $^{26}$Al-poor CAIs, as such a process would not introduce spatial heterogeneity on the scale of the protosolar disk (*11*). Hence, our solar system must have undergone either an in situ production or an external injection of SLRs shortly before the formation of the CAIs.

   A nearby supernova explosion has long been believed to be a strong candidate for the source of SLRs (*12, 13*). However, the supernova injection scenario faces an unresolved problem in that existing supernova models could not reproduce both the relative and absolute abundances of SLRs without disrupting the protosolar disk. For instance, these models predict that if a supernova provided $^{26}$Al and $^{41}$Ca to the solar system, it would also supply 100 times more $^{53}$Mn than its estimated nominal abundance (see Fig. 1(A) and Ref. (*14, 15*)). Moreover, regarding absolute abundances, Ref. (*16*) demonstrated that supernova explosions within 0.3 pc can disrupt the protosolar disk, and that a supernova injection event capable of supplying a sufficient SLR amount would likely prevent the solar-system formation altogether.

   To solve the discrepancy in relative abundance, an alternative combined scenario has been proposed in which $^{53}$Mn and $^{60}$Fe are injected from supernovae, while $^{26}$Al and $^{41}$Ca are synthesized by different processes (*15*), such as energetic particle irradiation from protosolar flares (*17*). This flare synthesis also could account for the presence of $^{10}$Be, which is absent in stellar nucleosynthesis and must originate from spallation reactions (*18,19*). However, even when adding in the contribution from the flare synthesis to the supernova injection (see **Materials and Methods**), the resulting SLR abundances differ by more than an order of magnitude from the nominal solar system values (see Fig. 1). It should also be noted that this flare synthesis has serious drawbacks in explaining the global distribution of SLRs in the solar system (*5*). Since the flare synthesis process works only in a minimal area of the protosolar disk, it would require extensive mixing on a scale not yet understood.

*Immersion model*

We propose a unified scenario, the immersion mechanism, that explains the origin of all SLRs to be consistent with the nominal abundances inferred from meteorites (Fig. 1). In this scenario, when certain SLRs—specifically $^{53}$Mn, and $^{60}$Fe—are injected into the protosolar disk from a nearby supernova, the disk is naturally immersed in accelerated particles confined within the shockwave of the supernova. This process can in principle drive in-situ synthesis of $^{10}$Be, $^{26}$Al, $^{36}$Cl, and $^{41}$Ca via non-thermal nucleosynthesis, a phenomenon we refer to as the immersion mechanism (see Fig. 2).

Our immersion mechanism assumes that a supernova explosion occurs in close proximity to the Sun ($d \lesssim 10$ pc) during the lifetime of the protosolar disk (see Birth Environment for the estimated event rate). In this model, the distance ($d$) plays a critical role in determining the supplied abundance of SLRs. When a supernova explodes, it generates a collisionless shock mediated by plasma instabilities, where charged particles, primarily protons, undergo diffusive shock acceleration to reach high energies (*20–22*). Theoretical studies on particle escape from supernova shockwaves indicate that the majority of accelerated particles, including sub-relativistic particles with energies of $\lesssim 1$ GeV, remain confined within the shocked region. Hereafter, these particles are referred to as 'trapped cosmic rays (trapped CRs)' (*23, 24*). As the supernova interacts with the protosolar disk, the heliosphere is compressed by the pressure of the supernova to a scale of $\lesssim 1$ au — smaller than Earth's orbital radius (*25–27*). This compression allows cosmic rays to impinge on the protosolar disk without hindrance from the magnetosphere. The disk could become largely exposed to trapped CRs, initiating non-thermal nucleosynthesis throughout its extent while simultaneously incorporating nuclides originating from the supernova, as in the injection model (*16, 28*). This nucleosynthesis process operates on gas and small grains in the disk. Given the stopping depth of $\lesssim 1$ GeV protons in rock (*29*), the trapped CRs can penetrate grains as large as 3–4 cm in radius. Thus, although the current model assumes that the supernova event happened before the formation of CAIs (typically smaller than this threshold (*30*)) as its baseline, this assumption can be relaxed.

To evaluate the predicted SLR abundances from our immersion mechanism, we formulated the ratio of the SLR abundance to the stable isotope (SI) abundance at the time of CAI formation, $N^{\text{SLR}}/N^{\text{SI}}$, under the assumption that pre-existing SLRs in the protosolar disk are negligible, as

$$\frac{N^{\text{SLR}}}{N^{\text{SI}}} = \frac{N^{\text{SLR}}_{\text{new}} \cdot \exp(-t_{\text{delay}}/\tau_{\text{SLR}})}{N^{\text{SI}}} \approx \frac{(N^{\text{SLR}}_{\text{inj}} + N^{\text{SLR}}_{\text{syn}}) \cdot \exp(-t_{\text{delay}}/\tau_{\text{SLR}})}{N^{\text{SI}}}, \quad (1)$$

where $N^{\text{SLR}}_{\text{inj}}$ and $N^{\text{SLR}}_{\text{syn}}$ denote the number densities of SLRs injected from the supernova and synthesized via trapped CRs, respectively. The sum of $N^{\text{SLR}}_{\text{inj}}$ and $N^{\text{SLR}}_{\text{syn}}$ is denoted by $N^{\text{SLR}}_{\text{new}}$, which represents the number density of SLR newly supplied to the protosolar disk. The exponential term represents the decay of SLRs with a mean lifetime $\tau_{\text{SLR}}$ over the time interval $t_{\text{delay}}$, which is the duration between the supply of new SLRs and the formation of CAIs. In Eq. (1), we assumed that injection and synthesis occur simultaneously. This assumption is justified because the time difference between these processes is expected to be much less than one million years (see **Materials and Methods**). The free parameters in this model are the time interval $t_{\text{delay}}$ and the distance $d$ between the supernova and the protosolar disk. The optimal model parameters are identified by minimizing the deviation between the predicted abundances of SLRs (see Eq. 1) and the nominal values derived from meteorite analyses (see Table 1).

For the non-thermal synthesis of SLRs in this system, we assumed that the disk is uniformly exposed to trapped CRs without temporal variation over the time interval $\Delta t$, which corresponds to the duration of the supernova shockwave traveling through the disk. The number density of SLRs, $N_{\text{syn}}^{\text{SLR}}$, synthesized by the bombardment of trapped CRs on a target nucleus labeled $j$ ($i + j \rightarrow$ SLR), can be expressed using the thin target approximation as

$$N_{\text{syn}}^{\text{SLR}} = \Delta t \sum_{(i,j)} \left( \gamma_i N_j \int_{E_0}^{\infty} \sigma_{ij}(E) \frac{dF_{\text{CR}}}{dE} dE \right), \tag{2}$$

where $N_j$ is the number density of the target nuclei $j$, $\gamma_i$ is the relative abundance of trapped CRs $i$ relative to protons, and $\sigma_{ij}(E)$ and $E_0$ represent the energy-dependent cross-section and the threshold energy of the reaction, respectively. These quantities are calculated using the TALYS code (*31,32*) (see **Materials and Methods** and Fig. 4). The number flux of accelerated particles in the supernova shock region, $F_{\text{CR}}$, is assumed to follow a standard power-law momentum distribution, $dF_{\text{CR}}/dE \propto p(E)^{-s}$, where $s \approx 2.1$ is the spectral index of CRs inferred from observations (*33*). The normalization of this flux is derived from a model that reproduces observed Galactic CR results (*34*), where 10% of the supernova kinetic energy density, $U_{\text{sh}}$, at the shock position is converted into CR energy density, $U_{\text{CR}}$. For the injection component, we adopted typical assumptions of the supernova injection model (*16, 28*). In this model, the supernova ejecta spread spherically, and only the fraction of SLRs intercepted by the protosolar disk is injected, depending on the distance $d$. We bracket uncertainties in CR acceleration efficiency, spectral index and progenitor mass in Fig. 5 (see **Materials and Methods** for more detail).

**Results**

Our immersion model offers a consistent explanation for the observed SLR abundances in the early solar system. Figure 1 shows that our immersion model successfully reproduces all SLR abundances in the early solar system to within one order of magnitude of their nominal values. This level of agreement falls well within the combined uncertainties, which stem from nuclear reaction cross sections and the estimated nominal SLR abundances. Each factor contributes to an overall uncertainty of approximately one order of magnitude. By contrast, each of the previously proposed models contains at least one SLR whose predicted abundance deviates from its nominal value by more than an order of magnitude. This discrepancy persists even when summing up the contributions of the flare and supernova injection. We also derived the optimal values of $d = 1$ pc and $t_{\text{delay}} = 0.45$ Myr, where $d$ is the distance from the supernova to the proto-solar disk and $t_{\text{delay}}$ is the time delay of CAI formation for the new SLR supply. Meteorite analyses constrain $t_{\text{delay}}$ to lie in the range of 0.2 Myr to 0.7 Myr (see **Materials and Methods** and Fig. 5(d)). Notably, the $^{41}$Ca/$^{40}$Ca ratio indicates that $t_{\text{delay}}$ should exceed 0.1 Myr, suggesting that immersion prior to CAI formation is desirable.

Beyond matching the SLR abundances, the immersion scenario addresses several limitations noted in previous scenarios. First, compared to the inheritance model from the molecular cloud, this mechanism can provide sufficient amounts of SLR while accounting for the nuclear decay that occurs during the time delay to CAI formation. Second, compared to the direct injection model by supernova explosions, this mechanism can reproduce both the relative and absolute amounts of SLR by a supernova at the distance that does not destroy the solar system ($d > 0.3$ pc) (*16*). Third, compared to the flare synthesis model, this mechanism does not require unexplained large-scale mixing in the disk, and can distribute SLR throughout the protosolar disk. Even the inner part of the protosolar disk can be exposed to trapped CRs, because the ram pressure of the supernova at 1 pc can compress the heliosphere to a scale of $\lesssim 0.1$ au (*26*).

The possible existence of a reservoir shielded from trapped CRs may help explain lower $^{26}$Al/$^{27}$Al ratios reported for a small fraction of CAIs (*35*). In our model, we assumed that the protosolar disk was optically thin to the CRs. However, the inner region of the disk could have been optically thick; based on the surface density profile of the minimum-mass solar nebula (*36*) and the cross-section of H$_2$ gas to ~1 GeV protons (*37*), the region within ~10 au is likely to have been opaque to CRs.

**Discussion**

*Birth Environment*

We find that typical young star-cluster environments readily permit the enrichment conditions required by our immersion model. Figure 3 shows that at least one supernova event occurs within 1 pc of the early solar system at high probability, if the Sun is formed within a star cluster (*38, 39*). The early solar system should have been in a young star cluster, and a large fraction of young star clusters are below the solid curve. Most observed clusters are located above the solid curve simply because they are old. They should have been below the solid curve when they were young (*40, 41*). This situation is further consistent with recent studies of star-forming regions, which depict environments characterized by dynamic interactions and frequent supernova explosions (*42, 43*). By contrast, the traditional supernova injection model requires a much closer explosion within 0.3 pc (*14,16*), which is statistically less probable. Only a small fraction of young star clusters are below the dashed curve.

Even if all the stars in a cluster formed during a few Myr, the clusters with total stellar masses ≳ 500 $M_\odot$ provide conditions under which a Sun-like star inevitably experiences a nearby supernova within its disk lifetime (*39*). Moreover, observations reveal that star formation in most clusters extends over 10 Myr or more (*44*), suggesting that the mass threshold is lower. Although a full, coupled model of disk survival, cluster dynamics, and supernova immersion is beyond our present scope, the joint evidence from cluster masses and age spreads offers statistical support for our scenario.

Once a massive star appears, its UV radiation clears residual molecular gas around the solar system (*42*), facilitating the direct injection of ejecta into the disk and efficient particle acceleration. Moreover, at distances ≳ 0.3 pc such UV flux remains too weak to photoevaporate the protosolar disk (*45*). Thus, its UV clearing would prepares the stage without destroying the target, leaving the immersion of the disk in ejecta physically plausible.

More than 50% of stars are born in massive star-forming regions comparable to or more massive than the Orion Nebula Cluster (ONC) (*46, 47*). Moreover, ~10% of stars still remain in bound clusters even after 30 Myr (*48*), corresponding to a timescale longer than the lifetime of massive stars that undergo supernovae. Thus, we conclude that at least 10%, possibly 50% of Sun-like stars are likely to host protoplanetary disks with SLR abundances similar to those of the protosolar disk.

*Universality of Our Solar System*

Our results suggest that Earth-like, water-poor rocky planets may be more prevalent in the Galaxy than previously thought, given that $^{26}$Al abundance plays a key role in regulating planetary water budgets (*1, 2*). Since a measurable fraction of stars form in clusters, solar-system-like SLR abundances are likely to be common rather than exceptional. This challenges previous interpretations that classified the solar system as an outlier with a particularly high $^{26}$Al abundance (*6*). Given our estimate that 10–50% of stellar systems in the Galaxy likely acquired solar-system-like SLR abundances with the immersion mechanism, we predict that upcoming exoplanet surveys targeting habitable zones around several tens of nearby solar-type stars, as proposed with the Habitable World Observatory (*49*), will lead to the detection of a few Earth-like rocky planets.

**Materials and Methods**
This study aims to reproduce the observed abundance ratios of short-lived radionuclides (SLRs) at the time of CAI formation. By treating the protosolar disk as a one-zone model, we investigated the optimal parameters that minimizes the deviation between the model-predicted values, as defined by Eq. (1), and the nominal $N^{SLR}/N^{SI}$ ratios derived from meteorite analyses (Table 1). As outlined in **Main Text**, the predicted values in our immersion model (Eq. 1) represent the sum of contributions from supernova injection and non-thermal nucleosynthesis. In the following sections, we address three key assumptions of our analytical model: (i) the timing of supernova injection and non-thermal nucleosynthesis, (ii) the reaction processes and cross-sections incorporated in the non-thermal nucleosynthesis term, (iii) the modeling of the supernova injection term.

*The timing of Injection and Immersion*
This section explains the rationale behind the assumption that supernova injection and non-thermal synthesis occur simultaneously in our model. This assumption is reflected in Eq. (1), where the two terms are added and assigned the same delay time. In practice, a time difference, $\Delta t$, exists between the moment the supernova ejecta make contact with the protosolar disk and the time the supernova shock region completely traverses the disk. At the point of contact ($d \sim R_{sh}$), the scale of the shocked region ($\Delta R_{sh} \sim 0.1$ pc) is considerably larger than the scale of the protosolar disk ($\sim 100$ au). Here, hydrodynamic simulations of supernova remnants show that the width of the shocked shell is approximately 10% of the shock radius $\Delta R_{sh} \approx R_{sh}/10$ (*50,51*). This configuration yields a timescale of $\Delta t \approx \Delta R_{sh}/v_{sh} \approx 43$ yr $(d/1 \text{ pc})(E_{exp}/10^{51} \text{ erg})^{-1/2}$, which is considerably shorter than the CAI formation timescale ($\sim 10^6$ years). Thus, the assumption in Eq. (1) is considered valid.

*Nuclear Reaction*
Here we provide a detailed account of the non-thermal nuclear reaction processes in our immersion model. All the reaction processes considered in this study and their respective cross-sections are presented in Fig. 4. The production of $^{60}$Fe by non-thermal synthesis was not considered in this study, and the cross-section for the production of $^{10}$Be was adopted from Ref. (*29*). The cross-sections for the production of other SLRs were calculated using the TALYS code (*31*).

The calculations presented in Eq. (2) do not encompass the full range of potential reaction processes. Instead, our analysis focuses on the interactions between protons/alpha particles with stable nuclei which have relatively high abundances, while excluding reactions with other CRs such as $^3$He particles. This exclusion is justified through the following estimation. Among the target stable nuclei $j$, the fraction $f$ converted to SLR by collisions with CR, labeled $i$, can be expressed as

$$f = \frac{N_{syn}^{SLR}}{N_j} \sim \langle \sigma_{ij} F_{CR} \rangle \cdot \Delta t. \qquad (3)$$

Using typical values for each parameter, the estimated fraction of stable isotopes depleted or synthesized in this nuclear reaction process is $f \sim 10^{-5} \gamma_i (\sigma_{ij}/100 \text{ mb})(d/1 \text{ pc})^{-2}$, where $\gamma_i$ is the relative number abundance of the CR nuclei $i$ compared to protons. This estimation suggests that the influence of the injected particles other than protons and alpha particles, as well as reactions involving low-abundance target nuclei and processes with small cross-sections, are negligible. This suggests that we can neglect the relative abundance $\gamma_i$ of impact particles other than protons and alpha particles, processes with a small cross-section $\sigma_{ij}$, and cases where the number density $N_j$ of parent nuclei is small by an order of magnitude.

It should be noted that the small value of $f \sim 10^{-5}$ indicates that the reduction in parent nuclei is sufficiently small, ensuring that any isotopic anomalies induced by our model remain minimal. This finding supports the robustness of our model in predicting the isotopic composition.

*Injection Model from Supernova*
This section evaluates the amount of SLRs injected into the protosolar disk from a nearby supernova, reviewing the material presented in Ref. (*16*). Supernova ejecta spreads spherically and is intercepted by the protosolar disk, which has a radius $R_{disk}$ and is located at a distance $d$. Hydrodynamical simulations have shown that the disk can resist complete destruction from supernova impacts at distances greater than $d > 0.3$ pc (*28, 52*). However, the contribution of gas-phase ejecta to SLR injection is minimal, accounting for less than 1% (*28*). Ref. (*53*) demonstrated that small dust grains ($\lesssim 0.1$ μm) follow the gas flow and are not injected into the disk, whereas larger grains ($\gtrsim 1$ μm) are injected with nearly 100% efficiency. Based on these findings, we assume the supernova ejecta consists of gas and dust, define the mass fraction of large dust grains ($\gtrsim 1$ μm) as $\eta_d$, and disregard SLR injection via gas and small dust.

Assuming that the injected SLR mass is uniformly mixed with the disk mass $M_{disk}$, the number density of injected SLRs, $N_{inj}^{SLR}$, can be expressed as

$$\frac{N_{inj}^{SLR}}{N^{SI}} \approx \frac{M_{inj}^{SLR}}{X^{SI} M_{disk}} \sim \eta_d \left(\frac{\pi R_{disk}^2}{4\pi d^2}\right) \frac{M_{SN}^{SLR}}{X^{SI} M_{disk}}, \quad (1)$$

where $M_{SN}^{SLR}$ represents the mass of SLRs ejected from the supernova, with a value adopted from Ref. (*54*), and $M_{disk} \gg M_{SN}^{SLR}$. The mass fraction of large dust grains is taken as $\eta_d = 20\%$, based on typical values observed in supernova SN 1987A (*55,56*). The timescale for SLRs to move from the supernova to the protosolar disk is approximately $t_{SN} \sim d/v_{SN} \sim 430$ yr $(d/1 \text{ pc})(E_{exp}/10^{51} \text{ erg})^{-1/2}$, suggesting that radioactive decay during transit is negligible.

*Combined Model*
Finally, we summarize the calculation method used to predict the values of the Combined Model, developed for comparison with our immersion model. To facilitate this comparison, Fig. 1 presents results from both our model and previous studies, including the Combined Model, which adds the contribution of the flare synthesis to the supernova injection. The development of the Combined Model was motivated by the recognition that both accelerated particle irradiation from the proto-sun flare and stellar nucleosynthesis likely contributed to the origin of SLRs in the early solar system. Despite this understanding, previous research lacked a quantitative model that integrated these processes. To address this gap, we derived the Combined Model as follows. The total amount of SLRs, $N^{SLR}$, in the Combined Model is given as the sum of SLRs supplied by flare synthesis and supernova injection, expressed as $N^{SLR} = N_{flare}^{SLR} + N_{inj}^{SLR}$. The flare synthesis contribution is based on Ref. (*29*), assuming a negligible delay time due to a nearly instantaneous synthesis. For the supernova injection term, we use the reference values from Ref. (*14*), denoted as $N_{inj,ref}^{SLR}$, and account for the distance $d$ using the relation $N_{inj}^{SLR}(d) = N_{inj,ref}^{SLR} \cdot (d/0.3 \text{ pc})^{-2}$. This term already includes a delay time of $t_{delay} = 0.4$ Myr, as specified in Ref. (*14*). To align this model with the nominal values of SLRs in the early solar system, we determined the distance d by minimizing deviations between the model predicted SLR abundances and their nominal abundances. The optimal distance for the Combined Model, shown in Fig. 1, is $d = 0.9$ pc.

*Model Uncertainties*

To quantify the robustness of our immersion model predictions, we have explored the sensitivity of the calculated SLR abundance ratios to key input parameters. In particular, we focus on two classes of uncertainty: (i) the properties of the CR population accelerated at the supernova remnant shock, and (ii) the choice of supernova yield models. Figure 5 illustrates the results of these tests.

The CR acceleration efficiency, $\epsilon_p$, and the spectral index, $s$, govern both the total energy injected into high-energy particles and the shape of their momentum distribution. We vary $\epsilon_p$ by a factor of four around our fiducial value of 10% (testing $\epsilon_p = 5\%$ and 20%; Fig. 5(a)), and vary $s$ around the nominal slope of 2.1 (testing $s = 2.0$ and 2.4; Fig. 5(b)). While lower acceleration efficiencies ($\epsilon_p = 5\%$) reduce the CR-synthesis contribution and diminish the ratios of $^{10}$Be, $^{26}$Al, $^{36}$Cl, and $^{41}$Ca by up to a factor of $\sim 2$, higher efficiencies ($\epsilon_p = 20\%$) enhance them by similar factors. Likewise, a harder spectrum ($s = 2.0$) depletes $\sim 100$MeV CR particles, which are most efficient at driving disk synthesis, and thus reduces CR-synthesis yields by up to a factor of $\sim 2$, as the effect of lowering $\epsilon_p$. Conversely, a softer spectrum ($s = 2.4$) enhances the low-energy CR flux and increases synthesis-derived isotopes by a similar factor.

Our baseline of the immersion model uses a 13 $M_\odot$ progenitor; however, nucleosynthesis yields depend on stellar mass. Supernova models alter the injected $^{53}$Mn and $^{60}$Fe yields via direct supernova–disk interaction, but their ability to reproduce meteoritic ratios varies with mass (Fig. 5(c)). While our fiducial 13 $M_\odot$ model provides the closest overall match, lighter progenitors (11–15 $M_\odot$) achieve comparably excellent agreement. This is an important result, since an initial mass function weighting favors these lower-mass stars. In contrast, the 25 $M_\odot$ and 40 $M_\odot$ models were adopted in previous studies *(14)* because they contain very high amounts of $^{26}$Al, but their feasibility has been questioned by recent observations *(57)*, and they may not represent typical supernova progenitors. Consequently, in our **Birth Environment** section, we restrict the supernova mass range to 8–20 $M_\odot$, ensuring both astrophysical realism and robust reproduction of the full SLR inventory.

Together, these sensitivity tests confirm that, despite factor-of-two variations from CR parameter choices and order-of-magnitude yield shifts from progenitor mass, our immersion model remains consistent with meteoritic SLR abundances within the adopted uncertainty envelope.

*Temporal Constraint on the Immersion Scenario*

To evaluate the temporal constraint of the immersion scenario, we varied the delay time $t_\text{delay}$ between the supernova encounter and CAI formation from 0.1 Myr to 0.9 Myr (Fig. 5(d)). $^{36}$Cl and, most notably, $^{41}$Ca—whose mean lifetime is only $\approx 0.99$ Myr—respond sensitively to this parameter. The meteoritic ratios remain within the acceptable uncertainty band for $0.2\,\text{Myr} \lesssim t_\text{delay} \lesssim 0.7\,\text{Myr}$, with an optimum around $t_\text{delay} = 0.45$ Myr. This time window corresponds to the epoch at which our immersion model can reproduce the standard $^{41}$Ca/$^{40}$Ca ratio. The 0.5 Myr of time window allowed for this model imposes physically reasonable constraints on the immersion process. At the same time, while immersion nucleosynthesis itself can occur after CAI formation, the requirement for $t_\text{delay} \gtrsim 0.1$ Myr suggests that immersion prior to CAI formation is preferable.

**Acknowledgments**

**Funding:**
JSPS KAKENHI Grant Numbers JP21K13964 and JP22KJ052 (R.S.).
Japan Science Society Grant Numbers 2025-2031 (R.S.)
JSPS KAKENHI Grant Numbers JP24H02236, JP24H02245, and JP24K00668 (Y.S.).
JSPS KAKENHI Grant Number 22H05150 (H.K.).
JSPS KAKENHI Grant Number JP21K13983 (T.T.).
JSPS KAKENHI Grant Number JP24K07092 (S.H.L.).
JSPS KAKENHI Grant Number JP24K07040 (A.T.).

**Figures and Tables**

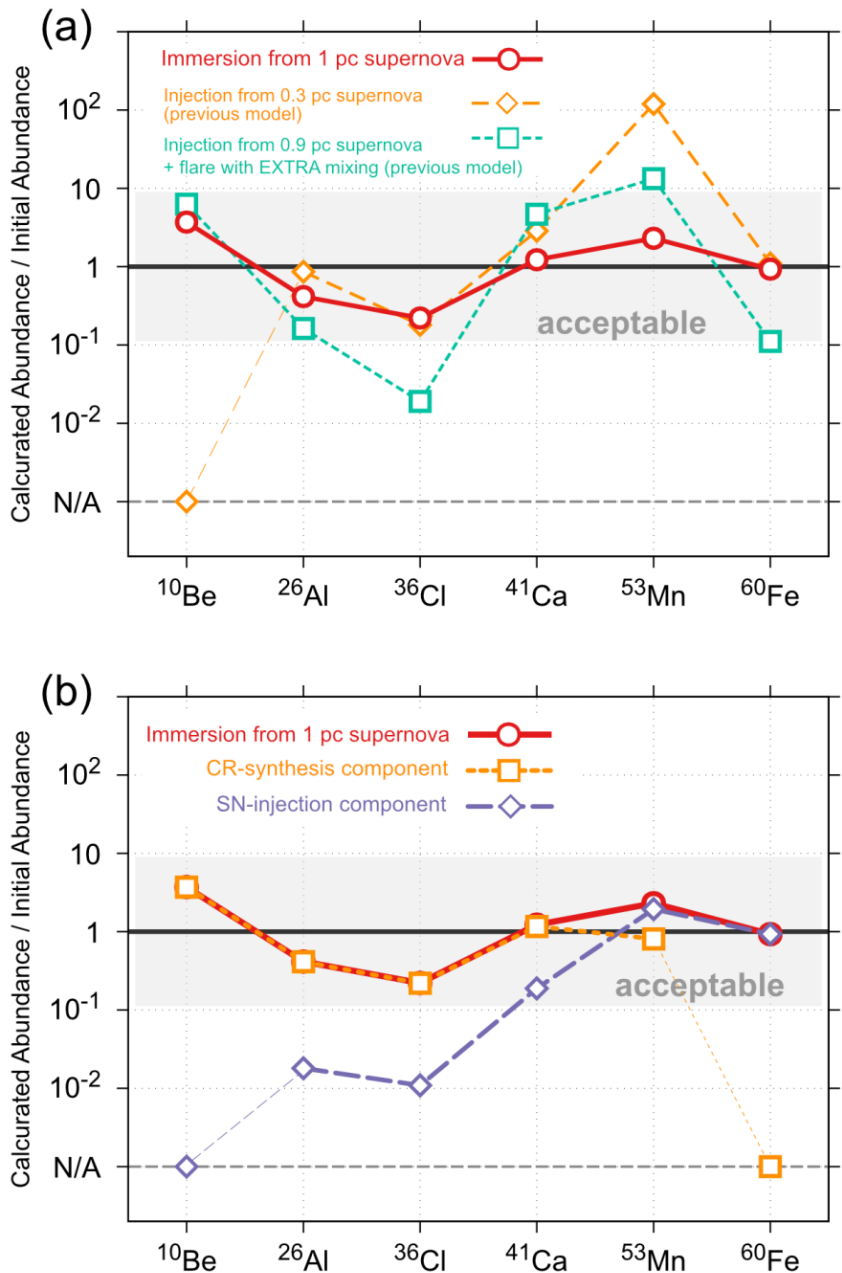

**Fig. 1. Normalized ratios (calculated/inferred nominal) for $^{10}$Be, $^{26}$Al, $^{36}$Cl, $^{41}$Ca, $^{53}$Mn, and $^{60}$Fe supplied by our immersion model, with the gray band indicating agreement with meteoritic constraints.** (**a**) Our immersion model (red circles; $d$ = 1 pc, $t_{delay}$ = 0.45 Myr, and 13 $M_\odot$ progenitor for optimal parameter) is directly compared to two other previous cases: only injection cases (orange diamonds; $d$ = 0.3 pc, $t_{delay}$ = 0.9 Myr, and 40 $M_\odot$ progenitor for optimal parameter) taken from Ref. (*14*), and the case of supernova injection + flare synthesis taken from Ref. (*29*) (see **Materials and Methods**, for more details). (**b**) Decomposition of the immersion result into CR-synthesis (orange dotted squares) and direct supernova injection (purple dashed diamonds) components, illustrating each individual contribution to the total SLR inventory.

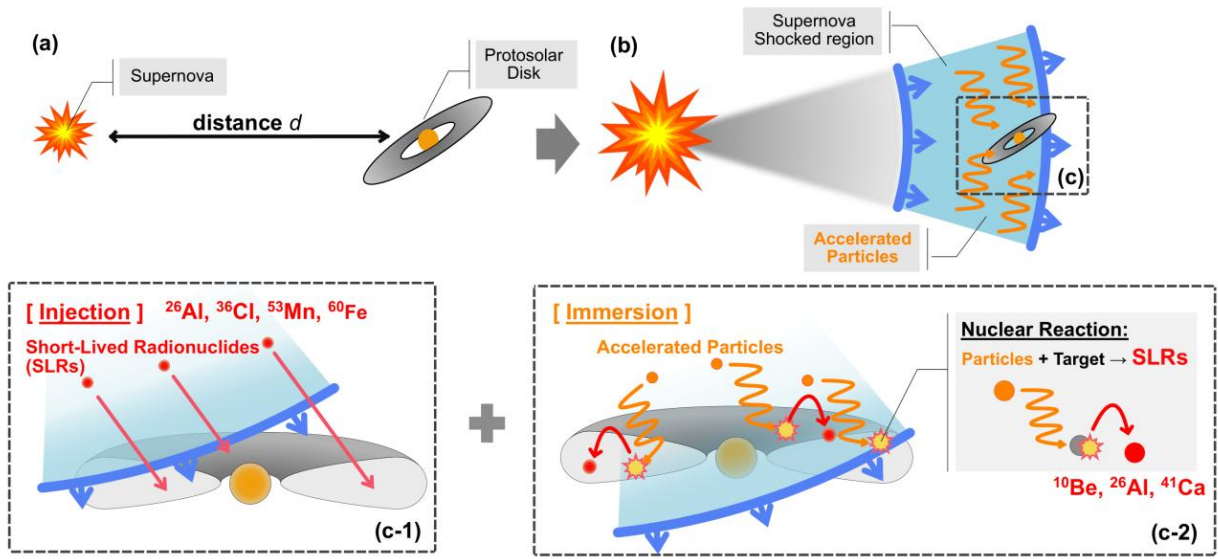

**Fig. 2. Schematic picture of the system assumed in this study.** (**a**) A supernova explosion occurs at a distance $d$ from the protosolar disk, and (**b**) the expanding supernova shockwave contacts the protosolar disk. At this time, a huge number of accelerated particles are trapped in the shockwave region. With the contact, (**c-1**) SLRs synthesized inside the supernova are directly injected into the disk, and (**c-2**) particles trapped inside the shockwave irradiate the disk, causing nucleosynthesis within the disk.

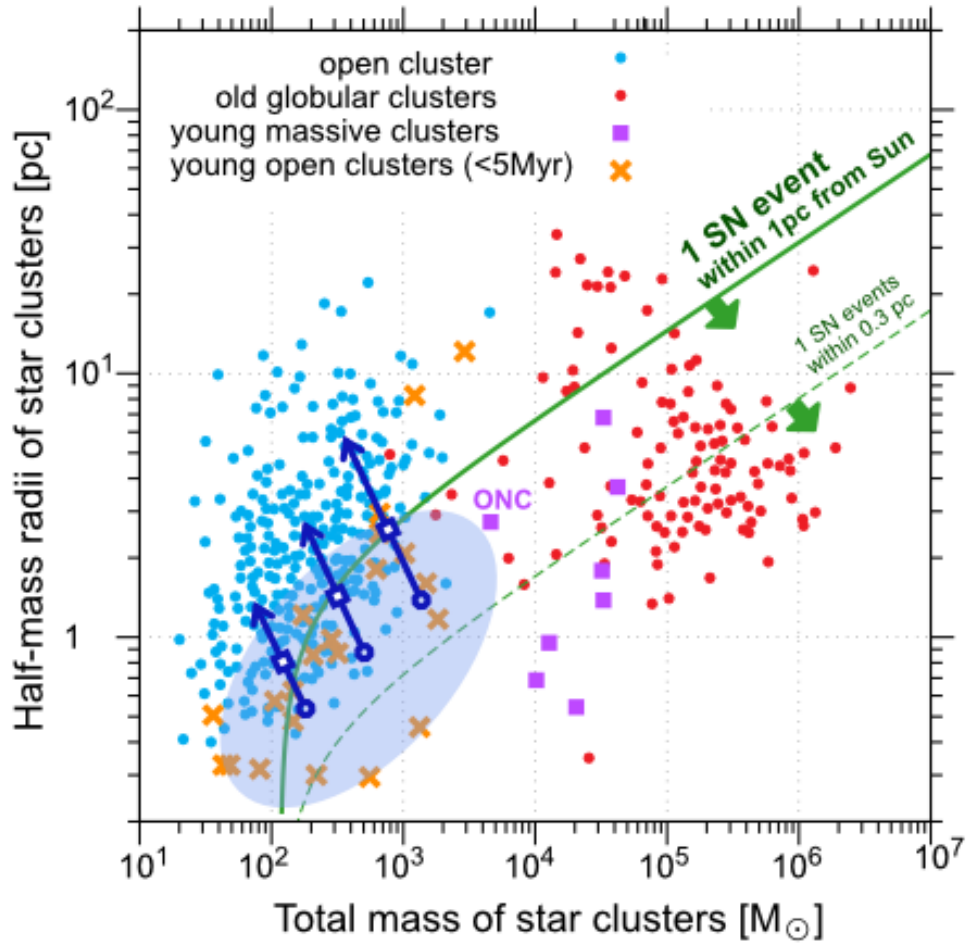

**Fig. 3: Diagram of half-mass radii and total masses of several types of star clusters: open clusters, old globular clusters, young massive clusters, and young open clusters.** A half-mass radius encloses half the mass of a star cluster. Data were obtained from Ref. (*48, 58*). The translucent blue region and the left-tilted arrows indicate the dynamical evolution of open clusters over time, as suggested by recent *N*-body+SPH simulations (*40, 41*). The open blue circle, open blue square, and arrowhead correspond approximately to $t = 0$ Myr, $t \sim 20$ Myr, and $t \gtrsim 20$ Myr, where $t$ denotes the age of the cluster since its formation. A star experiences at least one supernova event within 1 pc (0.3 pc) when it is in a star cluster below the solid (dashed) curve. Here, we modeled each star cluster as follows. Its stellar distribution follows the Plummer's distribution (*59*). We adopt a Kroupa initial mass function (*60*). Massive stars with $8 - 20\ M_\odot$ undergo supernova explosions (*57*).

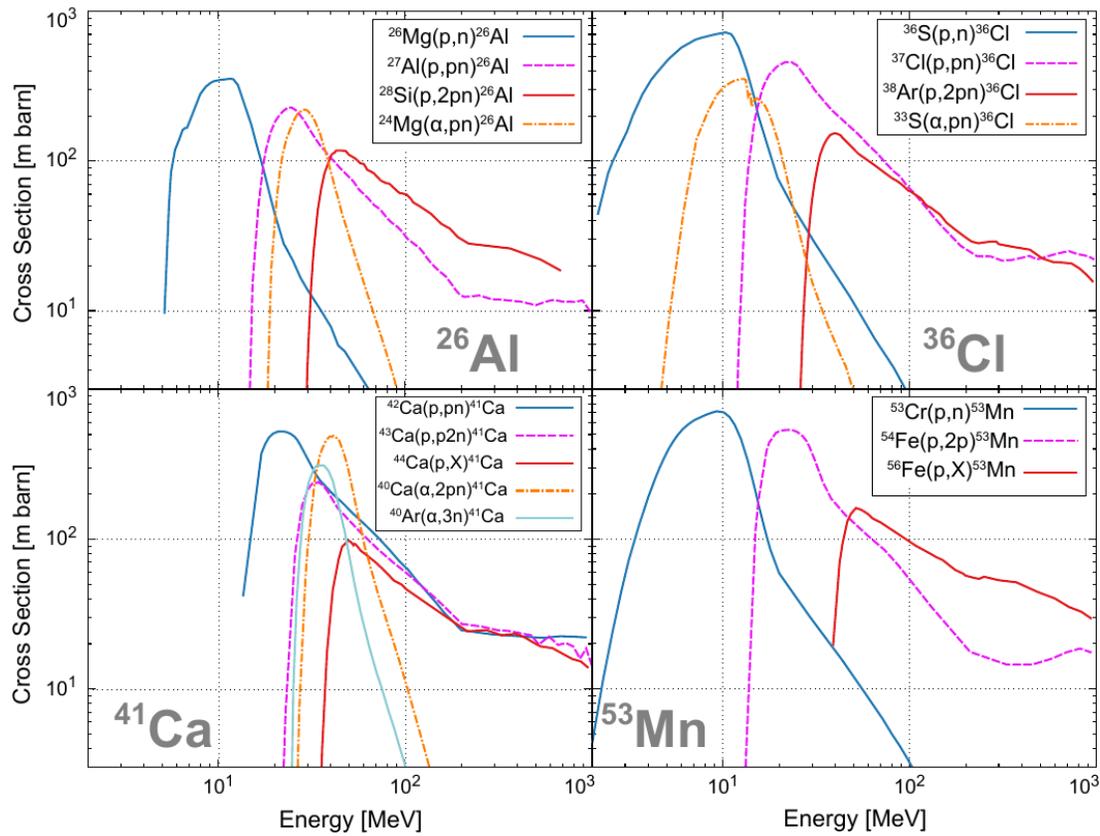

**Fig. 4. Energy-dependent nuclear reaction cross sections used in this study for the synthesis of $^{26}$Al, $^{36}$Cl, $^{41}$Ca, and $^{53}$Mn.** Each panel shows cross sections calculated by the TALYS code (*31*): the top-left panel for $^{26}$Al, the top-right for $^{36}$Cl, the bottom-left for $^{41}$Ca, and the bottom-right for $^{53}$Mn. For $^{10}$Be, we adopt the spallation processes $^{16}$O$(p,X)^{10}$Be and $^{16}$O$(\alpha,X)^{10}$Be, whose cross sections are taken from Ref. (*61*). No non-thermal production pathway is assumed for $^{60}$Fe under charged-particle collisions.

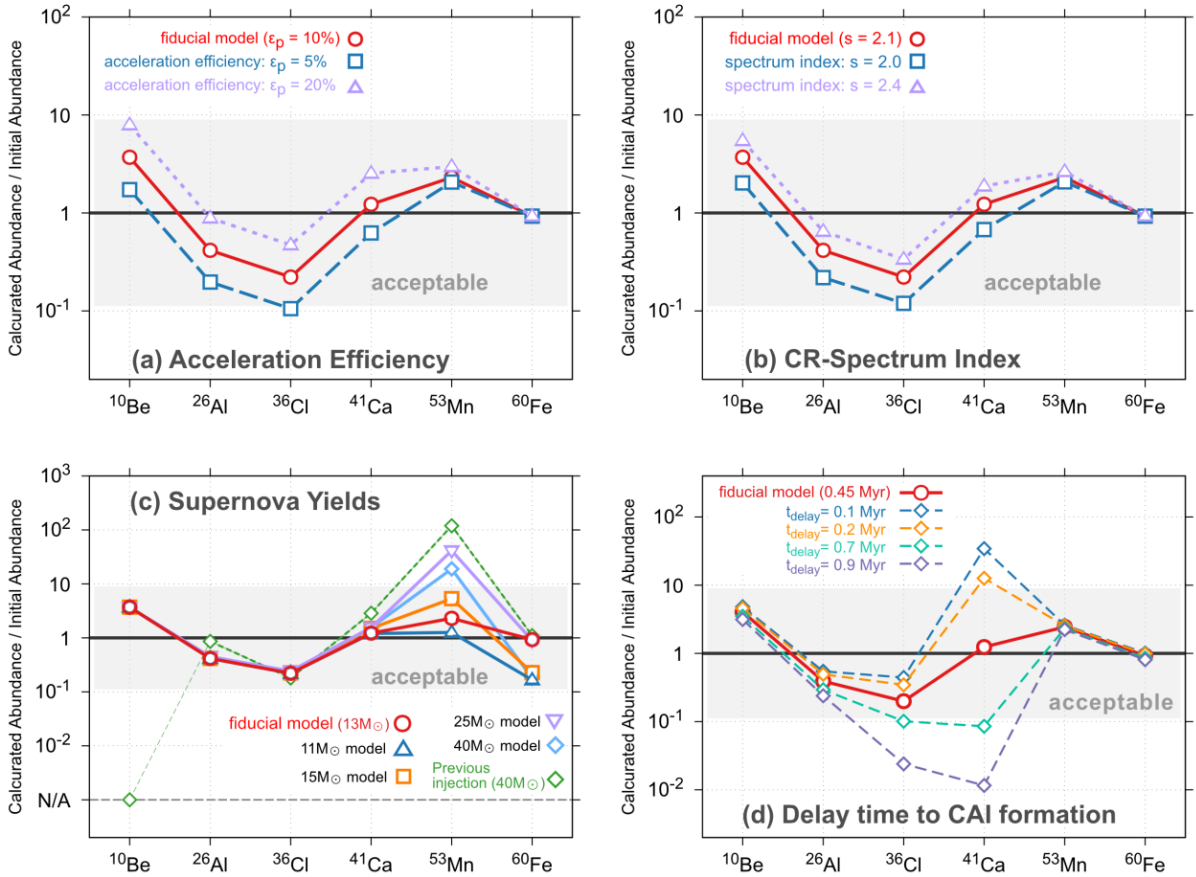

**Fig. 5. Uncertainty analysis for cosmic-ray parameters and supernova yields. Normalized ratios (calculated/inferred nominal) for $^{10}$Be, $^{26}$Al, $^{36}$Cl, $^{41}$Ca, $^{53}$Mn, and $^{60}$Fe supplied by our immersion model, with the gray band indicating agreement with meteoritic constraints.** (**a**) Sensitivity to cosmic-ray acceleration efficiency $\epsilon_p$: the fiducial value ($\epsilon_p = 10\%$, red circles) is compared to $\epsilon_p = 5\%$ (blue squares) and $\epsilon_p = 20\%$ (purple triangles). (**b**) Sensitivity to the cosmic-ray spectrum index $s$: the fiducial slope ($s = 2.1$, red circles) is compared to $s = 2.0$ (blue squares) and $s = 2.4$ (purple triangles). (**c**) Dependence on the adopted supernova progenitor mass: the fiducial case (13 $M_\odot$ model, red circles) is compared to yields from 11 $M_\odot$ (blue triangles), 15 $M_\odot$ (orange squares), 25 $M_\odot$ (violet inverted triangles) and 40 $M_\odot$ (cyan diamonds) models, alongside the prior work of Ref. (*14*) with 40 $M_\odot$ case (green diamonds). (**d**) Sensitivity to the delay time $t_{\text{delay}}$ between the super-nova encounter and CAI formation. The optimum model ($t_{\text{delay}} = 0.45$ Myr, red circles) is compared to $t_{\text{delay}}= 0.1$ Myr (blue diamonds), 0.2 Myr (orange diamonds), 0.7 Myr (green diamonds), and 0.9 Myr (purple diamonds). Only the short-lived isotopes ($\tau < 1$ Myr; $^{26}$Al, $^{36}$Cl, and especially $^{41}$Ca) show strong dependence.

**Table 1. Short-lived radionuclide (SLR) abundances in the early solar system and model predictions.** The evaluated values are expressed as the ratio of the number density of each SLR to its corresponding stable isotope ($N^{SLR}/N^{SI}$). The nominal values and half-lives of the listed SLRs were obtained from Ref. *(5)*. The last column presents the predicted values from our immersion model.

| SLR | Half-life (Myr) | Ratio | Nominal values | Our Model |
|---|---|---|---|---|
| $^{10}$Be | 1.387 | $^{10}$Be/$^{9}$Be | $7.1\times10^{-4}$ | $2.6\times10^{-3}$ |
| $^{26}$Al | 0.717 | $^{26}$Al/$^{27}$Al | $5.2\times10^{-5}$ | $2.2\times10^{-5}$ |
| $^{36}$Cl | 0.301 | $^{36}$Cl/$^{35}$Cl | $2.0\times10^{-5}$ | $4.5\times10^{-6}$ |
| $^{41}$Ca | 0.099 | $^{41}$Ca/$^{40}$Ca | $4.2\times10^{-9}$ | $6.3\times10^{-9}$ |
| $^{53}$Mn | 3.98 | $^{53}$Mn/$^{55}$Mn | $7.8\times10^{-6}$ | $2.1\times10^{-5}$ |
| $^{60}$Fe | 2.62 | $^{60}$Fe/$^{56}$Fe | $0.9\times10^{-8}$ | $1.0\times10^{-8}$ |

# Supplementary Materials for

## Cosmic-Ray Bath in a Past Supernova Gives Birth to Earth-Like Planets

Ryo Sawada *et al.*

*Corresponding author. Email: ryo@g.ecc.u-tokyo.ac.jp

**This PDF file includes:**

    Supplementary Text
    References (*62* to *73*) (if applicable—these should refer only to references in the SM)

## Supplementary Text

<u>Brief Summary of Previous Studies</u>
Three main hypothetical scenarios have been proposed for the origin of SLRs in the early solar system: inheritance, irradiation, and injection.

<u>Inheritance scenario</u>
  First, in the inheritance scenario, SLRs are already present in the protosolar molecular core before its collapse—either because they were passively mixed throughout the parent molecular cloud or because ejecta from a nearby supernova were injected into the parent molecular cloud locally. This definition thus covers both (i) passive inheritance from the natal cloud and (ii) injection of SLRs into the protosolar molecular core by a supernova shock that sweeps through the cloud after the bulk of the molecular gas has condensed.
  The main drawback of this scenario is the timescale problem: the time required for the solar system formation to begin is too long. SLRs with half-lives of less than 5 million years ($^{10}$Be, $^{26}$Al, $^{36}$Cl, $^{41}$Ca, $^{53}$Mn, and $^{60}$Fe) should decay to a negligible amount during the evolution from the molecular cloud core to the proto-Sun with the protosolar disk (*8, 9*). Ref. (*62*) recently proposed self-concentration of SLRs in a spiral-arm protostar cloud, but this still cannot deliver radionuclides with half-lives of less than 1 million years in time for CAI formation. Therefore, it seems necessary to provide SLRs in a way other than inheritance shortly before CAI formation.
  Beyond timing, the physical survivability of the injection of SLRs into the protosolar molecular core event is uncertain. Supernova remnants retain shock velocities ≥ 100 km s$^{-1}$ out to radii ≳ 20 pc (*63*), and simulations show that dense cores can be dispersed by shocks stronger than ∼ 70 km s$^{-1}$ (*64*). Therefore, the conditions under which SLRs could be injected locally into the parent molecular cloud while avoiding destruction at the distances proposed in the meteorite literature are not self-evidently satisfied and deserve quantitative testing beyond the scope of this study.
  Nevertheless, inheritance need not necessarily correspond to a single nearby event. Multiple supernovae at larger distances within the same or neighboring star-forming regions can provide a cloud-scale $^{26}$Al enrichment (*65*). We should note that this pathway moderates shock-destruction of dense cores, but relies on mixing efficiencies and transport times operating on longer spatial and temporal scales. In this sense, cloud-scale inheritance may have contributed modestly to the background level of SLRs in the protosolar environment, within which our immersion mechanism, that is, shorter timescale contamination from a nearby supernova, took place.
  Taken together, while cloud-scale enrichment cannot yet be ruled out, its temporal and spatial feasibility remains an open question.

Irradiation scenario

Next, the irradiation scenario proposes `in-situ' nucleosynthesis occurring in the protosolar disk due to the irradiation. $^{10}$Be is one remarkable nuclide that is produced only by the energetic-particle irradiation (e.g., Ref. (*29, 61*)) and cannot be supplied by other sources. The inferred value of $^{10}$Be/$^{9}$Be $\gtrsim 10^{-4}$ (*66, 67*) is too large to achieve by conventional Galactic Cosmic Ray irradiation (GCR-irradiation; (*29*)). Therefore, $^{10}$Be in the early solar system is mainly produced by the irradiation from a proto-Sun to a protosolar disk (PS-irradiation). Charged particles accelerated from the central star, especially during solar flares, possess sufficiently high energy (~ 10-100 MeV) to cause such nucleosynthesis (*68*). Classically, the so-called X-wind model (*69*), which includes nucleosynthesis processes due to the PS-irradiation for rocky vapors, has been proposed and widely accepted. However, numerous concerns were pointed out with the X-wind model (*70*), and it is considered that the X-wind model could not explain the solar system either from a nucleosynthesis perspective. Now Ref. (*17*) is the only viable and quantitative model, including the PS-irradiation. This model still faces a problem from the viewpoint of the homogeneity of the SLRs within the protosolar disk. Since the PS irradiation occurs in the minimal region near the central proto-Sun, it is required that an efficient transport mechanism mixes materials throughout the protosolar disk (*68*).

Injection scenario

Finally, the injection scenario suggests the possibility that the SLRs are directly injected into a protosolar system from a nearby core-collapse supernova explosion, as the lifetime of a massive star is short (less than 10 million years for stars with a mass of 20 $M_\odot$ or more (*71*)). The initial abundance of $^{60}$Fe, which is exactly the opposite of $^{10}$Be, is the key to unveiling the origin of SLRs. This is because $^{60}$Fe is barely synthesized by energetic-particle irradiation around the young Sun (e.g., Ref. (*61*)). In the injection scenario, the difficulty lies in the conditions under which direct injection can be achieved without destroying the solar system. There are two possibilities: direct injection into the parental molecular cloud core (*72*) and into the protosolar disk (*28*). Ref. (*73*) argued that the chromium isotopic heterogeneity found in various chondritic meteorites might be evidence of the direct injection onto the protosolar disk. However, Ref. (*16*) ruled out direct injection into the protosolar disk as the only explanation for the origin of SLRs, such as an insufficient amount of $^{26}$Al.